\documentclass[showpacs,showkeys,pre,amssymb]{revtex4}
\usepackage{amsmath}
\usepackage{bm}
\usepackage{multirow} 
\usepackage{graphicx}

\usepackage{setspace}
\newcommand{\spps}[1]{_{\scriptscriptstyle{#1}}}
\newcommand{\bdsy}[1]{\boldsymbol{#1}}

\begin{document}

\title{Influence of coaxial cylinders on vortex breakdown in a closed flow.}

\author{Cecilia Cabeza$^{a}$}
\author{Gustavo Saras\'ua$^{a}$ }
\author{Arturo~C.~Mart\'{\i}$^{a}$ }
\author{Italo Bove$^{a}$ } 
\author{Sylvana Varela$^{a,b}$}
\author{Gabriel Usera$^{b,c}$}
\author{Anton Vernet$^{b}$}

\affiliation{$^{a}$Instituto de F\'{\i}sica, Universidad
de la Rep\'ublica, Montevideo, Uruguay}

\affiliation{$^{b}$Departament d'Enginyeria Mec\'anica, Universitat Rovira i Virgili,
Tarragona, Spain}

\affiliation{$^{c}$Instituto de Mec\'anica de los Fluidos, Facultad de
Ingenier\'{\i}a, Universidad de la Rep\'ublica, Montevideo, Uruguay}

\date{\today}


\begin{abstract}
The effect of fixed cylindrical rods located at the centerline axis on vortex breakdown (VB) is  studied experimentally and numerically.  We find that the VB is enhanced for very small values of the rod radius $d$, while it is suppressed for values of $d$ beyond a critical value. In order to characterize this effect, the critical Reynolds number for the appearance of vortex breakdown as a function of the radius of the fixed rods and the different aspect ratios was accurately determined, using digital particle image velocimetry. The numerical and experimental results are compared showing an excellent agreement. In addition, a simple model in order to show that this effect also appears in open pipe flows
is presented in the appendix.

\end{abstract}


\maketitle

\section{Introduction}
\label{sec:int}The development of structural changes in
vortical flows and, particularly, vortex breakdown
\cite{Peck,Vogel1968,Leib78,Leib84,Escudier1984}
has been intensively investigated during the last years
\cite{Lopez1990a,Lopez1990b,Lopez95,Lopez96,Mark2003,Mullin2000,Mitchell2001,
Husain2003,Mununga2004,Fujimura2004,Piva2005,Lim2005,Zhang2006}. The
characteristic and fundamental signature of vortex breakdown is the
appearance of a stagnation point followed by regions of reversed axial
flows with a bubble structure when the swirl is sufficiently large.
This structural change is also accompanied by a sudden change of the
size of the core and the appearance of disturbances downstream the
enlargement of the core. Vortex breakdown is very important in several
applications of Fluid Mechanics such as aerodynamics, combustion,
nuclear fusion reactors or bioreactors. The presence of this phenomenon 
in these devices may be beneficial or detrimental, depending on each
particular application \cite{Husain2003,Mununga2004}.

Vortex breakdown (VB) has been firstly observed in open flows \cite{Peck}, but
subsequently it has been obtained also in experiments performed in confined flows, 
for example, in closed cylinders \cite{Escudier1984}. It is worth noting that the
characteristics of  VB in both cases are strongly similar,
suggesting the possibility that the basic mechanism of VB is the same
in both situations \cite{Mark2003}. Experimental
measurements show that flows in open channels can be accurately
described with the relatively simple q-vortex model \cite{Leib84}. On
the other hand, an analytical description for closed flows which are
considerable more complex than closed flows is not known
\cite{Escudier1984,Lopez1990a,Lopez1990b}. Thus, experiments performed with flows confined
in cylinders are very attractive in order to understand the basic mechanism of VB, as well as 
investigate possible control mechanisms.

Due to the great number of practical implications of VB, the development of mechanisms for controlling its emergence is of
considerable interest. Recently, different methods for controlling VB in closed flows have been proposed using different techniques. 
The effect of adding a small, rotating or stationary, cylinder was first investigated
using numerical and experimental techniques, by Mullin {\it et al}  \cite{Mullin2000}.
Husain {\it et al}  \cite{Husain2003} considered  the addition of co- or counter rotation near the axis using a small central
rod rotating independently of the bottom or rotating end wall.  They concluded that
such method may be an effective way to either enhance or suppress VB.
Other approaches are based on the introduction of a small rotating disk in the end wall opposite to the rotating wall \cite{Mununga2004,Yu2006,Yu2007}, or the use of axial temperature gradients \cite{Herrada2003}. 

In the Ref.~\cite{Mullin2000}, the authors found that a straight small cylinder does not change qualitatively
the characteristics of the flow.  They also considered the case of a stationary cylinder for fixed aspect ratio and Reynolds number
they did not find important qualitative changes with respect to the case without inner cylinder.
They concluded that it is necessary to slope the inner boundary in order to significantly alter the VB bubble.
However, as these authors recognize it is difficult to find a reliable measure for the onset of VB using dye visualization. 
It is clear that a more accurate technique is necessary in order to get quantitative experimental results for the VB onset.

The aim of this work is to analyze accurately the effect of the presence of a rod 
at the cylinder axis on the VB for an important range of parameters. Since the onset of VB is associated
with the appearance of a stagnation point near the axis an accurate determination
of the velocity field is necessary. Using the Digital Particle Image Velocimetry
(DPIV) technique we obtain the critical Reynolds number for the onset of VB as a function of the aspect ratio and radius of
the inner cylinder. In order to gain a deeper understanding of the internal structure of the flow and accuracy in 
the critical Reynolds numbers, we considered necessary to complement our experimental observations with numerical simulations.
It is found that the existence of rods can affect significantly the critical value of the Reynolds number, $R_c$ for the onset of the VB. 

This paper is organized as follows.  In Sec.~\ref{experiment} we
present the experimental setup and general observations. In section
III, we describe the numerical method used in the simulations. The
comparison between the experiments and the numerical calculations is given in Sec.~\ref{resultados}. Finally, in Sec.~\ref{summary}, we present a summary and the conclusions.

\section{Experimental Setup}
\label{experiment} 

\begin{figure}
\begin{center}
\includegraphics[width=.30\columnwidth]{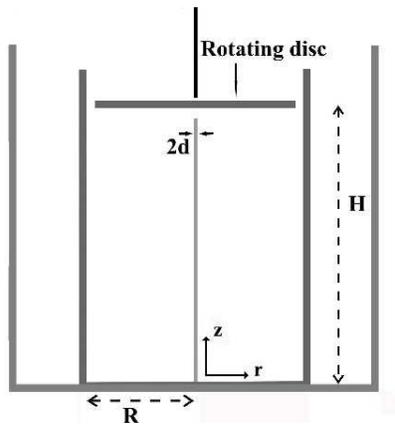}
\end{center}
\caption{The experimental setup consists of a closed cylinder, radius $R$ and height
$H$, with a rotating top wall. Along the axis of the cylinder a fixed
rod of radius $d$ is located. A external container was used in order to minimze the optical distortion of the images due to the curvature of th
e cylindrical wall.}
\label{setup}
\end{figure}

The experimental setup consists of an plexiglas cylindrical container
of inner radius $R = 40.0$ mm, a rotating disk at the top and a fixed
disk at the bottom of the cylinder (see figure \ref{setup}). The rotating disk is driven by an electronically controlled DC motor,
operating between $\Omega=1$ rad/s and $\Omega=21$ rad/s, with an error of less than $0.5$\% in the measurment of $\Omega$. The height $H$ of t
he flow domain, and therefore the aspect ratio $H/R$, can be varied by moving the rotating disk up and down to a predetermined location.  In this work,  four different aspect ratios $H/R$: 1, 1.5, 2 and 2.5 were used.

In a first series of experiments, the emergence of vortex breakdown in absence of rod
for comparison with results of previous works is studied.
In the second part of the experimental work,  the effect of axial rods on VB development is investigated.
We use the cylinder described above, with an axial rod fixed on the bottom disk. The rod was
perfectly aligned along the axis of the cylinder and was kept at rest
during the experience. Three different rod sizes, with
corresponding radius $d=1.0$ mm, $d=2.5$ mm and $ d =5.0$ mm were used.

The fluid used was water dissolutions of glycerin at 60\% in mass,
with $\nu = 1.0 \times 10^{-5}$ m$^2$/s. Neutrally buoyant polyamide particles of $50\times10^{-6} \, m$ diameter was seeding inside fluid in o
rder to obtain the velocity profile. A vertical green laser sheet
of 100 mW, with $2$ mm of
width was used as an illumination system. The sheet was focalized at the meridional plane of the
cylinder. The images of the flow were record using a CCD camera at 90
fps. In order to minimize optical distortion of images due to the
curvature of the cylindriIn the onset of VB a stagnation point in the axis appears in the flow . In this case, the vertical velocity at the axi
s is null and we used this fact to

\begin{figure}
\begin{center}
\includegraphics[width=.40\columnwidth]{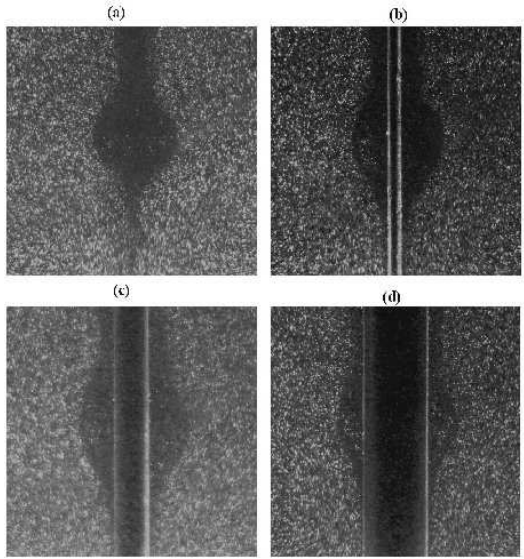}
\end{center}
\caption{Magnifications of the zone corresponding to the recirculation bubble at the onset of VB for different radius of the rod.
The aspect ratio ($H/R=2$) is the same in all the images while the $Re_c$ varies in each image: (a)
$\mathrm{Re}=1474$, without rod.  (b) $\mathrm{Re}= 1313$, $d=1.0$mm, (c)
 $\mathrm{Re}=1431$, $d=2.5$mm. (d)$\mathrm{Re}=1671$, $d=5.0$mm.}
\label{completo}
\end{figure}

Figure~\ref{completo} shows the typical DPIV images for the onset of VB corresponding to different radius of the rod.cal wall, the container 
was immersed in a rectangular box filled with the same working fluid, since both the
solution and the plexiglas have similar refractive indices.
The experiments were conducted in an air-conditioned room at $20^o$C
and the fluid temperature was monitored regularly using a thermocouple
located at the bottom of the external container. The Reynolds number
was defined as in previous studies as $Re=\Omega R^2/ \nu$, which
varies in the experiments between 600 and 2900, with a 2\% of error.

The conventional Digital Particle Velocimetry Velocity (DPIV) technique was implemented in order to obtain the velocity field inside the flow.
This technique allow us to get accurate quantitative values of the velocity profile based on the cross-correlation of two consecutive images. I
n fact, this technique allow us deteminate accurately the critical Reynolds number ($Re_c$) at the onset of VB through the measurment of the vertical velocity at the axis of the cylinder. The onset of VB is typically associated with localized regions of reverse flows, i.e. an internal stagnation point on the vortex axis appears, located at a certain height, ($z_c$). The appearance of VB bubbles is a manifestation of internal separation of the streamlines from the axis, causing the reversal of the axial velocity. We used the criteria that VB arises when the axial velocity changes the sign at the axis. In the situation when a fixed rod is localizate at the axis of the cylinder, we analyze the vertical velocity at the wall of the rod.

Figure~\ref{completo} shows the typical DPIV images for the onset of VB corresponding to different rod radius.

\begin{figure}
\begin{center}
\includegraphics[width=16cm]{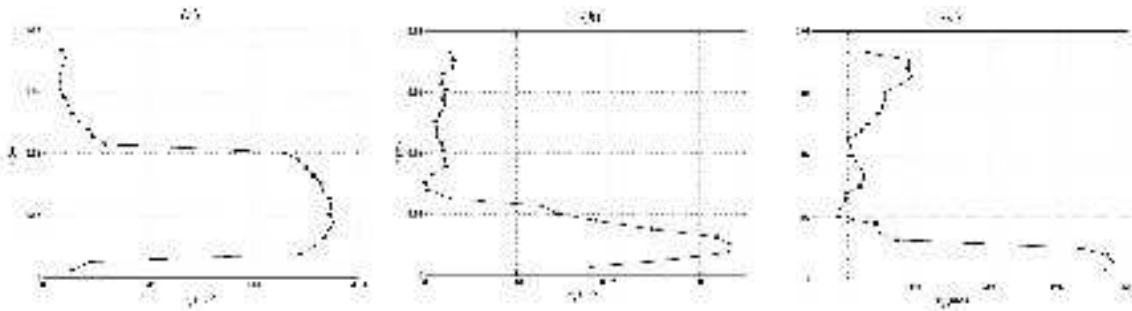}
\end{center}
\caption{
Axial velocity profiles at the axis ($r=0$) for $H/R=2$ obtained using the DPIV technique and corresponding to the cylinder without rod .
(a) situation before the onset of VB, $\mathrm{Re}=1391$, the axial velocity at the axis is allways positive. (b) first VB, $\mathrm{Re}=1474$, the axial velocity at the axis present a null value at the stagnation point located at $\mathrm{z_c}=3.1$ cm. (c)  onset of the second VB, $\mathrm{Re}=1836$, in this case the axial velocity at the axis presents a range of negative values in the place of the first reciculation bubble and a second stagnation point at  $\mathrm{z_c}=4.5$ cm where the second bubble is going to appear.}
\label{vz}
\end{figure}

In figure ~\ref{vz} three typical axial velocity profiles at the axis ($r=0$) for $H/R=2$ correspondig to the cylinder whitout rod are show. In Fig. \ref{vz} (a) previous to the onset of VB the axial velocity is
always positive and in this case the Reynold number is $Re=1391$. At the onset of VB, Fig.~\ref{vz} (b) corresponding to $Re_c=1474$, the axial velocity present a null value at the stagnation point located at $\mathrm{z_c}=3.1$ cm above the bottom of the cylinder. Finally, in Fig.~\ref{vz} (c) we can observe that the second VB take place at $Re_c=1836$. In this case the axial velocity at the axis presents a range of negative values in the place of the first reciculation bubble and a second stagnation point at $\mathrm{z_c}=4.5$ cm above the bottom of the cylinder.

\section{Numerical Method}
\label{simula}

The numerical simulations considered here were obtained with the
in-house flow solver caffa3d.MB developed jointly by Universitat Rovira i Virgili (Tarragona, Spain) and Universidad de la Rep\'ublica
(Montevideo, Uruguay). It is an original Fortran95 implementation of a fully implicit finite volume method for solving the 3D incompressible
Navier-Stokes equations in complex geometry, using block structured body fitted grids.  This three-dimensional solver is based on a
two-dimensional solver described in \cite{Ferziger02}. Further description of this model can be found in Ref.~\cite{usera06,codigo}.

The mathematical model comprises the mass (\ref{eq:MassBalance}) and
momentum (\ref{eq:uMomentumBalance}) balance equations for an
incompressible Newtonian fluid:
%
\begin{equation}
\label{eq:MassBalance}
\int_{S}
\left(\vec{\bdsy{v}}\cdot\hat{\bdsy{n}}\spps{S}\right)dS = 0
\end{equation}
%
\begin{displaymath}
\int_{\Omega} \rho\frac{\partial u}{\partial t}d\Omega + 
\int_{S} \rho u
\left(\vec{\bdsy{v}}\cdot\hat{\bdsy{n}}\spps{S}\right)dS=
\end{displaymath}
\begin{equation}
\label{eq:uMomentumBalance}
\int_{S} -p\hat{\bdsy{n}}\spps{S}\cdot\hat{\bdsy{e}}\spps{1}dS+
\int_{S}
\left(2\mu\bdsy{D}\cdot\hat{\bdsy{n}}\spps{S}\right)
\cdot\hat{\bdsy{e}}\spps{1}dS
\end{equation}

These equations hold in any portion $\Omega$ of the domain, being S
the boundary of $\Omega$ and $\hat{\bdsy{n}}\spps{S}$ the outward
normal vector at the boundary S.  The momentum balance equation
(\ref{eq:uMomentumBalance}) has been expressed for the first component
$u$ of the velocity vector $\vec{\bdsy{v}}=(u,v,w)$, with similar
expressions holding for the other components. The viscosity $\mu$ of
the fluid and the symmetric deformation tensor $\bdsy{D}$ were used
for the viscous term.

Since the Reynolds number was relatively low for all cases, since
$\mathrm{Re}=800$ to $\mathrm{Re}=2700$, no turbulence model was
required so transient solutions were computed directly. The time step
was set to $10^{-1}$ s for all cases. Simulations were run starting
from null velocity fields through $10^3$ time steps, or about 100 s of
flow time.

\begin{figure}
\begin{center}
\includegraphics[width=.60\columnwidth]{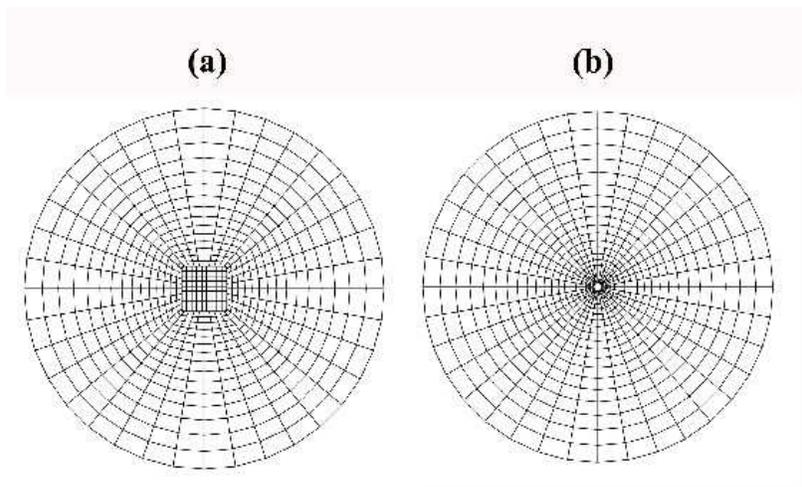}
\end{center}
\caption{
(a) Cross section of the block-structured grid, with a central square block, used in the case of
the cylinder without rod. This grid topology prevents the need for internal boundary conditions at the
axis of the domain for the case without rod.
(b) Cross section of the structured grid for the case of the cylinder with rod.}
\label{f9}
\end{figure}

Two different grids were used, one to study the fluid without rods,
and other to study the system including the rods. To study the system
with rod  C-grid block, with a no-slip boundary condition at walls, Fig. \ref{f9}(b), is used. 
The different choice of grid topology for the two cases reflects that a boundary
condition near the axis exists for the case that includes rods, while no boundary
condition exists at the axis for the case without rods. Thus, the two-block grid strategy
adopted for the second case ensures that the critical region near the axis is treated
entirely as an inner region in the absence of a rod, not requiring the specification of
delicate  boundary conditions there. On the other hand, for the case that includes rods,
the single cylindrical block is the natural choice allowing a direct specification of the non--
slip boundary condition at the rod.
On the other hand, the grid used for the cylinder without rods is a block structured grid, made of
two blocks. The inner one is a prism of a square base of $5$ mm
($0.2592\times 10^{6}$ cells) located in the axis of the cylinder,
while the outer one is a C-grid block that wraps around the first one,
see Fig. \ref{f9} (a) ($2.304\times 10^{6}$cells).

\begin{figure}
\begin{center}
\includegraphics[width=.6\columnwidth]{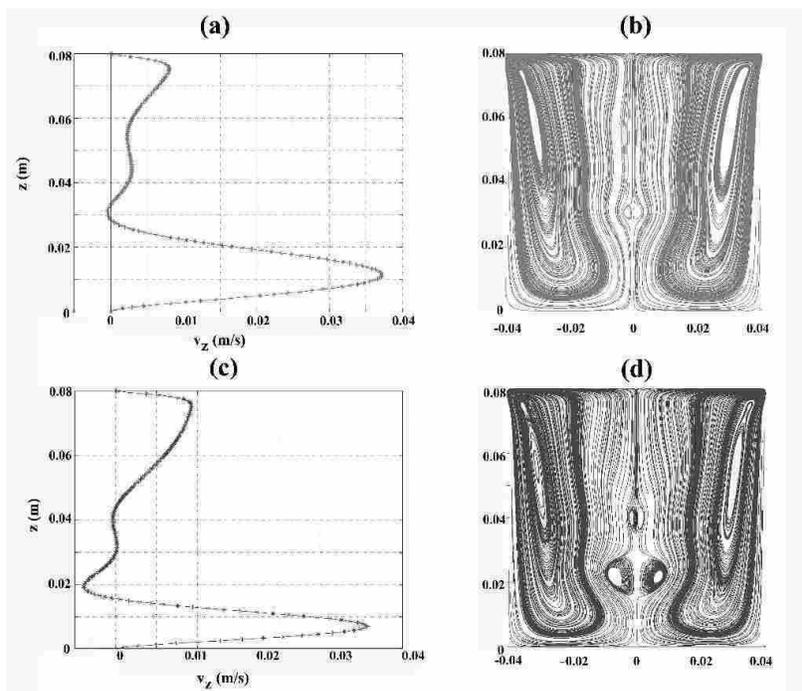}
\end{center}
\caption{
Numerical results of the axial velocity profiles (left) and streamlines (right) corresponding to the same situation of figure ~\ref{vz}.
(a) and (b) one bubble is observed for $H/R=2$ and $\mathrm{Re}=1474$. (c) and (d) two bubbles are observed for  $H/R=2$ and $\mathrm{Re}=1836$.
 }
\label{burbusimula}  
\end{figure}

The numerical model allows us to get the three components of the
velocity field ($v_r$, $v_\theta$, $v_z$) and its derivatives. As
mentioned before, the reversals of the axial component of the velocity
$v_z$ in the region near the axis gives us information on the
emergence of areas of recirculation (bubbles), which are a way of
determining VB. In Fig.~\ref{burbusimula} numerical results of the
axial component of the velocity are showed for the cylinder without
rod. The two showed panels correspond to one (top) and two (bottom)
VB bubbles. This velocity profiles show an excellent agreement with
those obtained experimentally (Fig.~\ref{vz}).

The simulations are performed for different cases (three values of the
aspect ratio $H/R$, cylinder with and without rods) and Reynolds
numbers. The numerical results are in excellent agreement with the
experimental velocity profiles. The position of the stagnation
point is obtained with a difference less than $5\%$ in the system with and without
rods. The numerical results were analyzed trough streamlines $\psi$ and
azimuthal vorticity $\eta_{\theta}=\eta_y\cos\theta-
\eta_x\sin\theta$. In addition, the energy density, defined as
$\emph{e} = \frac{p} {\rho} + \frac {1}{2}(v^2_r + v^2_{\theta} +v^2_z)$, 
the velocity module and the pressure field were obtained (not shown in this work).

\begin{figure}
\begin{center}
\includegraphics[width=.90\columnwidth]{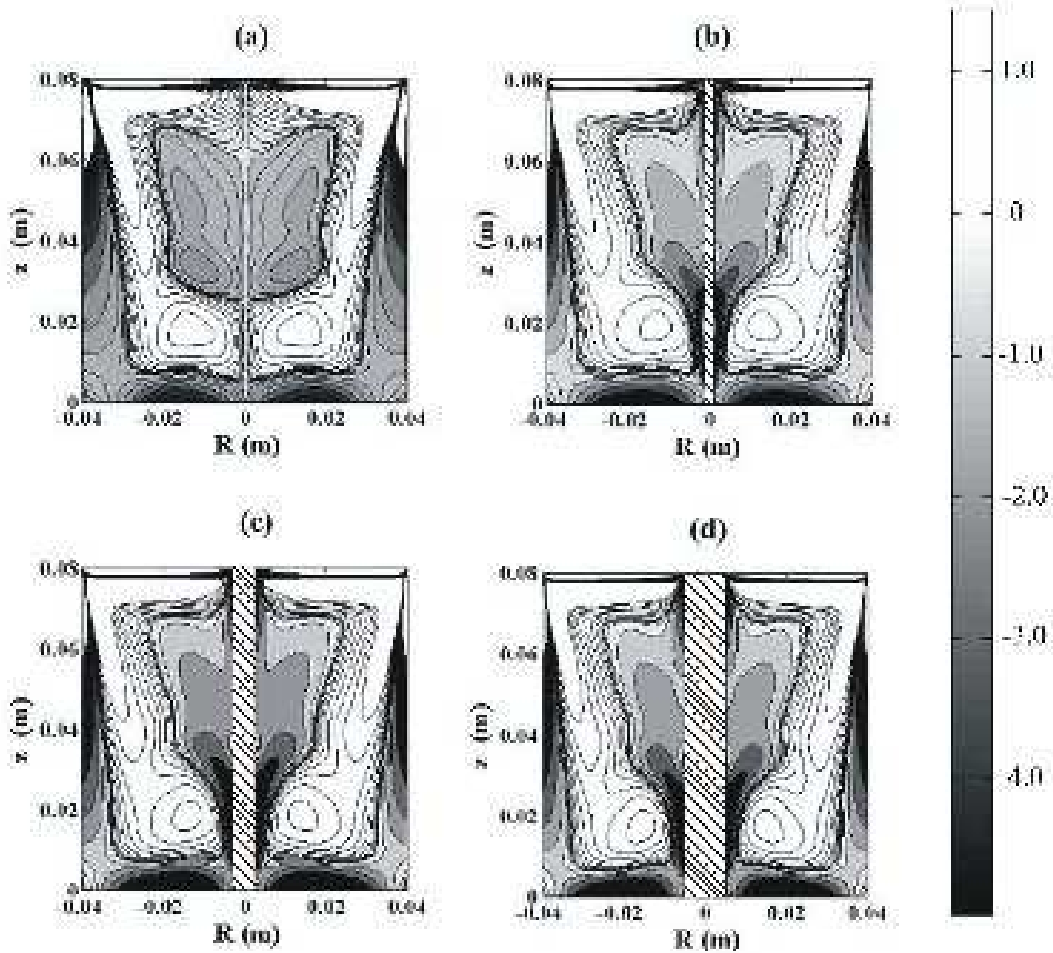}
\end{center}
\caption{
Numerical results for the azimuthal component of the vorticity at the onset of VB and aspect ratio $H/R=2$. The lined regions represent the fixed rod. (a) without rod, $\mathrm{Re}=1474$, (b) fixed rod of radius $d=1.0$ mm and  $\mathrm{Re}=1340$. (c) fixed rod of radius $d=2.5$ mm and $\mathrm{Re}=1435$.  (d) fixed rod of radius $d=5.0$ mm and  $\mathrm{Re}=1700$. The scale, indicated in the colorbar at the right, is the same
for all the panels.
}
\label{vort1}
\end{figure}

Figure \ref{vort1} shows contours for the  azimuthal vorticity
$\eta_{\theta}$ in the meridional plane for $H/R=2$
at  critical Reynolds number $\mathrm{Re}_{c}$ as indicated. 
The contour levels are non-uniformly spaced, with 20 levels in the
 positive range and 20 in the
negative one  determined by
$level_{pos}(i)=Max(variable)\times(\frac{i}{20})^3$ and
$level_{neg}(i)=Min(variable)\times(\frac{i}{20})^3$
\cite{Lopez1990a}.  All are plotted at $t=100$ s, when steady-state
flow conditions have been reached.

\section{Results and Discussion}
\label{resultados}

For convenient comparison with previous works \cite{Escudier1984}, we
first studied the emergence of vortex breakdown in absence of rods.
In table ~\ref{tab1} the experimental and numerical results for the onsent of VB for different aspect ratios are shown. In this table also Escudier results are shown. First, if we compare our experimental and numerical results, we can observe that its present an excellent agreement with a difference lower than 2\%.  In second place, our results are in very good agreement with the results of Escudier \cite{Escudier1984}. Our results present a difference lower than 3\% respect to Escudier's results. We can conclude that the method experimental and mumerical that we have used are very reliable. Thus, the DPIV technique provide an accuracy method to determinate the onset of the VB through the analize of vertical velocity profile. The result presents in this work shown that the DPIV technique are more realible that one based in visualization using dye technique.

Now, we consider the situation in which an axial fixed rod of radius $d$ is present. In table ~\ref{tab1} for $d/R= 0.025$ we can obaserve that the Reynolds number necessary to achive the onset of VB is less than the case without fixed rod. Meanwhile for $d/R=0.0625$ the onset is achived for $Re_c$ slightly lower than $d/R=0$ and for $d/R=0.1250$ the $Re_c$ present values 15\% above than the case without rod.

\begin{figure}
\begin{center}
\includegraphics[width=.90\columnwidth]{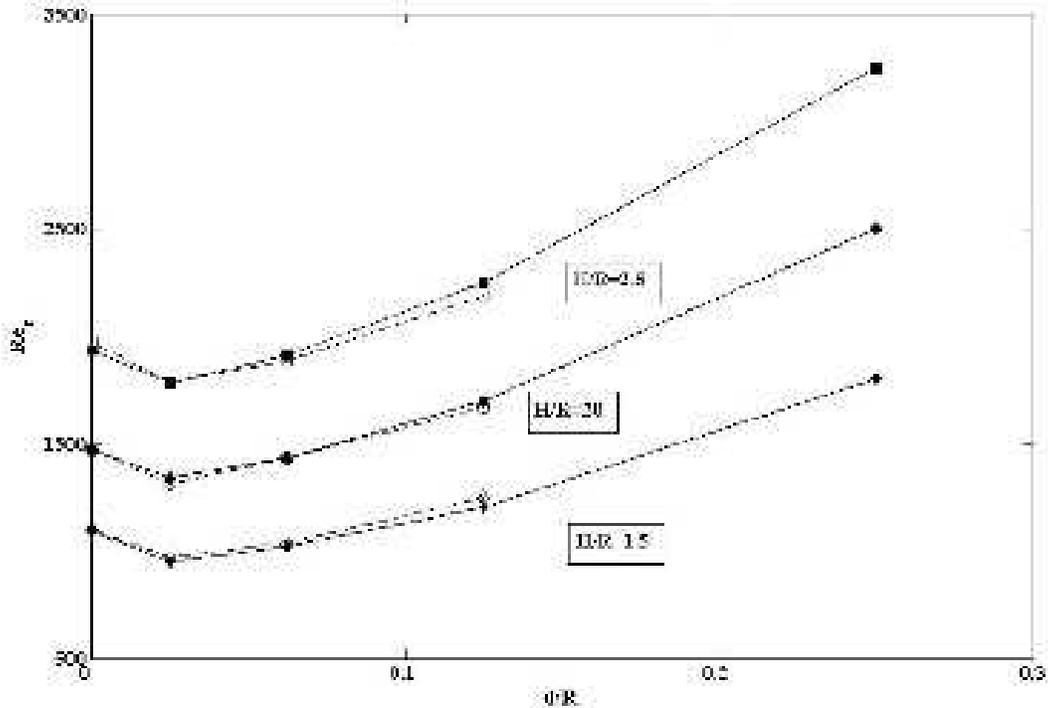}
\end{center}
\caption{Experimental (open symbols) and numerical (filled symbols) values of
the critical Reynolds numbers corresponding to the appearance of the first bubble as a function of the rod radius, for different values of the
aspect ratio $H/R =1.5$ (diamonds), $H/R =2.0$ (circles) and $H/R =2.5$ (squares)
}
\label{diag}
\end{figure}

In figure Fig \ref{diag} is depicted a comparison between the
experimental and numerical results concerning the effect of the cylinders on the
first bubble formation, showing that there is a very good agreement. We show the $Re_c$ as function $d/R$ for the three aspect ratio that we have studied. In order to analyze the response of the system when the radius of rod is increased, we calculated numerically the $Re_c$ for $d/R=0.25$. The difference in the Reynolds number at the onset is approximately 65\% between cylinder without and with rod, for all aspect ratio analyzed here. In summary, we can conclude that if the aspect ratio is kept constant, increased the radius of the rod represent a significative change in the $Re_c$ at the onset and for small radius of the rod, the $Re_c$ is lower than the case  without rod. However, there is a region of radius of the rod, that depend of aspect ratio, which slightly variation is observed for the $Re_c$ between cylinder with and without rod.

A clear modification is noted also in the vorticity structure, as can be seen
from Fig.~\ref{vort1}.  For $d = 2.5$ mm and $d = 5.0$ mm, we observe that the critical Reynolds number to begin vortex breakdown is
increased. The effect in the structure of the vorticity in these cases
are similar to the observed for the smallest rod.  Experimental and
numerical for the critical value $Re_c$ are summarized in Table~\ref{tab1}.

\begin{table}
\begin{center}
\begin{tabular}{|c||c|c|c|c|c|c|c|c|c|}  \hline
H/R &\multicolumn{3}{|c|}{d/R=0}&\multicolumn{2}{|c|}{d/R=0.0250}&\multicolumn{2}{|c|}{d/R=0.0625 }&
\multicolumn{2}{|c|}{d/R=0.1250} \\  \hline
      & Exp.~(1). & Exp.~(2)   & Num.~& Exp.~(2)    & Num.~ & Exp.~(2)     & Num.~ & Exp.~(2)  & Num.~  \\  \hline
  1.5 & 1068  &1098   & 1095& 971     & 950 & 1036      & 1020& 1246   & 1200 \\  \hline
    2 & 1450  &1474   & 1474& 1313    & 1340& 1431      & 1435& 1671   & 1700 \\  \hline
  2.5 & 1910  &1973   & 1935& 1785    & 1780& 1887      & 1910& 2187   & 2250 \\  \hline
\end{tabular}
\caption{ Experimental and numerical critical Reynolds numbers corresponding to the appearance
of the first VB  for different aspect ratios and radius of the rod. The experimental values (1) were taken
from \cite{Escudier1984} and the (2) were obtained in this work using the DPIV technique.}
\end{center}
\label{tab1}
\end{table}

It is interesting to compare our results with those obtained in
references \cite{Mullin2000}, \cite{Husain2003}. In these two works,
it was concluded that a very slender rod at rest does not introduce
significant changes in VB phenomena. The radius ratio used by these
authors was $d/R=0.04$, which is similar to the small radius ratio
value we considered, that is $d/R=0.025$.  In reference
\cite{Mullin2000}, the authors used an inner cylinder with ratio
$d/R=0.1$.  In our experiments we observed appreciable changes in the
critical value of $\mathrm{Re}$, for all values of the ratios
considered.  Looking carefully at Figures 2 and 3 of
\cite{Mullin2000}, we observe slight differences which, according to
the authors, are within the experimental errors. For this reason, we
took specially careful measurements in these cases.

At this point it is interesting to consider the origin of the effect of the rod on VB,
and in particular if it is of viscous nature or not. Inviscid theory has largely used to
explain the origin of VB. The results has been sufficiently succesfull to support the
belief that  VB may occur with no presence of viscosity. In ref. \cite{Lopez1990b},
Brown and Lopez argumented that VB in closed flows takes place in conditions for which
the flow may be considered as inviscid. In this case, a  change of sign in the azimuthal
component of vorticity $\eta_{\theta}$ is expected to occurs  near the stagnation point \cite{Lopez1990b}. This condition is verified
in the VB  that takes place without rod, as can be seen in figure \ref{vort1}a .
In this  figure, the stagnation point occur at $z = 0.032$, where $z$ measures the
height from below.  However, this condition  is not satisfied in the presence of the rods. In these cases,
the stagnation point occur near $z = 0.03$ on rod surface. This point is surrounded with a region of constant
$\eta_{\theta}$ sign, as can be seen in figures \ref{vort1}b,c,d. Thus the relation of Brown and Lopez does not holds, showing that we cannot expect to describe the VB appropriately with only inviscid  theory in the presence of rods.

The decrease of $Re_c$ due to small rods appears to be of viscous
origin.  The presence of the rods tends to reduce the velocity of the
fluid due to stress at the rod surface, favoring the appearance of the
stagnation point and thus reducing the critical value $Re_c$ needed
for the appearance of VB.  The reducing effect of the viscosity is
also present in the absence of rods and may begin the process that
leads to the formation of the stagnation point (as occur in pipe flows
\cite{Lopez1990b}), but the presence of the rigid limit imposed by the
rod enhances this effect very much. Thus, a very thin rod produces an
significative effect, because the boundary conditions at the axis
changes drastically from free to no-slip.  This produces an
appreciable change in $Re_c$.  A similar result was obtained in
previous works where it has been shown that even though a thin wire may promote
the formation VB in delta wings \cite{Akilli}.  On the other hand, we
note that when the diameter $d$ is sufficiently large, the critical
value of $Re$ increases with increasing $d$ (see Fig. \ref{diag}).
This effect is probably of inviscid origin. We support this conjecture
with the results of a simple model for flows in a pipe, described in
the appendix.  With this model we obtain that the suppressing effect
of increasing $d$ also appears in unconfined pipe flows. Then, when $d$
increases, this effect of inviscid nature compensates the viscous one
and finally the VB is displaced to higher values of Re.

\section{Summary and Conclusions}
\label{summary}

In this work, we studied the changes in the VB phenomena that are produced when 
a rod 
is located at the axis of the cylinder container i.e. the vortex centerline.
The experiments  shown
that the rod may increase or decrease  the critical Reynolds number for the
emergence of VB.
The effect of small rods is to decrease $Re_c$, while  $R_c$ is increased 
when $d/R$ is beyond a certain critical value $(d/R)_c$,
which is of order $0.06$. This critical value depends weakly on the
 aspect ratio, decreasing smoothly with increasing $H/R$. 
The effects of the rods of practical interest since may be used to perform a control 
of the VB  in a way that 
is simpler than other previously
proposed in the literature, since it does not require additional
auxiliary devices. It is worth noting that this method, unlike the
approach proposed in previous works, does not imply
the addition of swirl near the axis.  This simplicity is 
 an interesting feature,  making it more feasible to be used in
engineering devices. The modification of the critical Re are very 
noticeable. The volume ratio (cylinder to row) is $V_1/V_2
\sim 4 \times 10 ^{-3}$, while the decrease of the critical Reynolds
number is about $10\%$ . So that the shift of the critical Reynolds
number is 20 times larger that the percent modification of the volume
of the cylinder, showing the effectiveness of the method. The dependence of
 the critical Re with the rod diameter obtained experimentally is in very good 
agreement with the found with  numerical calculations.
Using a simple theoretical model for flow in pipes, we obtained that the influence
 of the inner cylinder on the VB works in similar fashion in open flows, mainly for
not too small values of $(d/R)_c$. We gave arguments to support that the  enhancing 
effect of very small rods on VB is of viscous origin.

We acknowledge financial support from the Programa de Desarrollo
de Ciencias B\'asicas (PEDECIBA, Uruguay) and Grants FCE 9028 
and PDT54/037 (Conicyt, Uruguay).

\begin{center}
{\small \bf APPENDIX}
\end{center}

\label{model}
In this appendix, we consider a model for  VB in
open swirling pipe flows. Our interest is to study how  the VB is affected
when an inner cylinder is located at the axis of the duct.
With this purpose, we  use a simple model based on the failure of 
the cylindrical solution. 
In the past, it has been shown that simple theoretical vortex
models of open flows may predict with some accuracy the emergence of
 vortex breakdown. A classical approach based on
axisymmetric inviscid analysis was given by Batchelor
\cite{Batch}.  In this model, it is assumed that initially the
fluid moves inside a cylindrical duct of radius $a_{1}$\ , with a Rankine
vortex velocity field, and goes to another duct limited by two cylindrical boundaries 
of different radius. When the flow develops the stagnation point, 
the radial component of the velocity becomes important and the cylindrical approximation
is no longer valid \cite{Leib78}. This is reflected in the disappearance of cylindrical 
solutions.
Thus in this model, the disappearance of cylindrical solutions is identified with the
emergence of VB \cite{Batch}.

In a cylindrical system of coordinates
($r,{\theta},z$)
 the velocity field of the assumed upstream flow is written as:  
\begin{eqnarray}
\label{f1}
\ v_{r}&=&0, \\
v_{\theta }& =&\begin{cases}
             \sigma r  & 0 < r < a_2, \\
              \Gamma /r & a_{2}<r<a_{1},
	      \end{cases}  \\
 v_{z}&=&U_{0}.
\end{eqnarray}
where $\sigma ,\Gamma$ and $U_{0}$ are constants, with $\Gamma =\sigma
a_{2}^{2} $, in order to impose the continuity of the velocity. We introduce the 
 streamfunction $\Psi$ defined as
\begin{eqnarray}
 v_{z}=\frac1r \frac{\partial \Psi}{\partial r}, \  \ 
v_{r}= - \frac 1r \frac{\partial \Psi}{\partial z}.
\end{eqnarray}
Employing the momentum conservation we obtain the following
equation for the streamfunction:
\begin{equation}
r\frac{\partial }{\partial r}(\frac{1}{r}\frac{\partial \Psi
}{\partial r})+ \frac{\partial ^{2}\Psi }{\partial
z^{2}}=r^{2}\frac{de}{d\Psi }-K\frac{dK}{ d\Psi }  \label{eqphi}
\end{equation}
where $e=\frac{1}{2}v^{2}+\frac{p}{\rho }$ and $K=rv_{\theta }$.
For the flow given by Eqs.~(\ref{f1}), we have that
\begin{equation}
e=\frac{2\sigma ^{2}}{U_{0}}\Psi
+\frac{1}{2}U_{0}^{2},\;\;K=\frac{2\sigma }{ U_{0}}\Psi ,\; for \;
0<r<a_{1}  \label{relphi}
\end{equation}

The same relations hold downstream for the steady flow, and so that
Eq. (\ref{eqphi}) takes the form for the rotational part, 
if we restrict ourselves to consider 
cylindrical solutions,

\begin{equation}
\Psi (r)=\frac{1}{2}U_{0}r^{2}+AF_{1}(\gamma r)+BY_{1}(\gamma r)
\label{solphi}
\end{equation}
where \ $F_{1}$\ and $Y_{1}$ are the Bessel functions of the first and
second kind respectively and $\gamma = 2 \sigma / U_0$ \cite{Batch}. The
irrotational part of the flow is given by
$ v_{\theta }=\Gamma /r, \; v_{z}=U, \; v_{r}=0 $.

We particularize now to consider the situation in which the upstream flow goes two 
a duct limited by to cylindrical 
surfaces of inner radius $d$ and outer radius $R$. 
The unknown constants which appear in Eq. (\ref{solphi}) can be
determined imposing the mass conservation and the continuity of
the pressure. From these conditions, we obtain the following
equation that gives the value of the radius of the rotational core 
in the downstream duct $c$

\begin{equation}
\frac{(a_{1}^{2}-a_{2}^{2})}{(R^{2}-b^{2})}-\frac{A_{II}}{U_{0}}%
\gamma J_{0}(\gamma b)-\frac{B_{II}}{U_{0}}\gamma Y_{0}(\gamma
c)-1=0,
\end{equation}
with
\begin{equation}
A_{II}=\frac{U_{0}}{2 c d}[\frac{d(a_{2}^{2}-c^{2})Y_{1}(%
\gamma d)+c d^{2}Y_{1}(\gamma b)}{J_{1}(\gamma
c)Y_{1}(\gamma d)-J_{1}(\gamma d)Y_{1}(\gamma
c)}],
\end{equation}
\begin{equation}
B_{II}=\frac{U_{0}}{2 c d }[\frac{d (a_{2}^{2}-c^{2})J_{1}(%
\gamma d)+c d^{2}J_{1}(\gamma b)}{Y_{1}(\gamma
c)J_{1}(\gamma d)-Y_{1}(\gamma d)J_{1}(\gamma
c)}].
\end{equation}

\begin{figure}
\begin{center}
\includegraphics[width=.60\columnwidth]{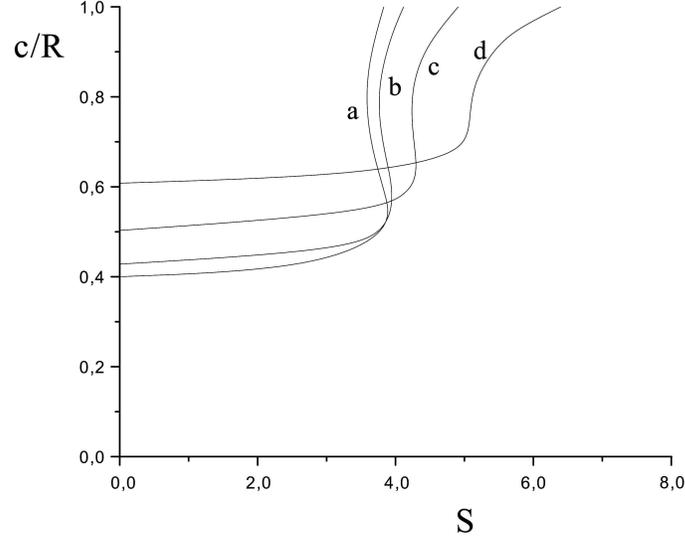}
 \end{center}
\caption{Dimensionless core size, $c / R$,
as a function of the swirl, $S$, for different radius of the fixed
rod, (a) $d=0$, (b) $d=0.2$, (c) $d=0.4$, (d) $d=0.6$ and (e) ,
$d=0.8$. Other parameter values: $R_0=1$, $c_0=0.4$ and $R=1.2$}
\label{ratio2}
\end{figure}

\begin{figure}
\begin{center}
\includegraphics[width=.60\columnwidth]{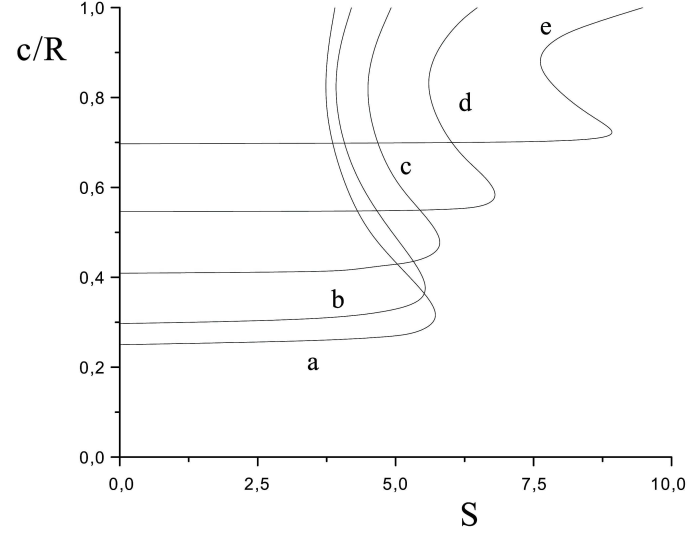}
 \end{center}
\caption{As figure ~\ref{ratio2} but with $R_0=1$, $c_0=0.25$ and
$R=1.2$, for (a) $d=0$, (b) $d=0.2$, (c) $d=0.4$, (d)
$d=0.6$.}
\label{ratio1}
\end{figure}

In figures \ref{ratio2} and \ref{ratio1} it is shown the value of $c$
 as a function of the nondimensional swirl parameter $S$, defined as
 $S = \gamma R$. From these figures we can see that at certain
 critical values $S_c$ of $S$, two branches of solutions collide and
 disappear. According to the failure of the cylindrical approximation
 criteria, this is the value of $S$ for which VB takes place.  From
 these figures, we can see that the effect of a very slender inner
 cylinder may be to increase or diminish slightly the critical value
 $S_c$ (the first case is observed in figure \ref{ratio2} and the
 second in figure \ref{ratio1}). However, for all the situations
 considered, $S_c$ increases with $d$, as long $d$ is above a
 threshold.  In this case, the emergence of VB was transferred to
 higher values of the nondimensional swirl parameter $S$ and the VB
 was suppressed in the range of $S$ values contained between the old
 and new critical values of $S_c$.  These results shows that the
 present inviscid model predicts the suppressing effect of the slender
 cylinders, in analogy with the experimental observations of the
 closed flow, showing that viscosity is not necessary to explain this
 behavior.  However, the decreasing effect on $S_c$ that is observed
 for small values of $d$ in figure \ref{ratio1}, has nothing to do
 with the observed in the closed flow for the smallest of value of
 $d$, where the effect appears to be of viscous origin.

\end{document}